\documentclass[journal,comsoc]{IEEEtran}
\usepackage[T1]{fontenc}
\usepackage[english]{babel}
\usepackage[utf8x]{inputenc}
\usepackage{url,enumerate}
\usepackage{graphicx}
\usepackage{amsmath}
\usepackage{cite}
\usepackage{amssymb}
\usepackage{lineno}
\usepackage{comment}
\usepackage{subcaption}
\usepackage{dblfloatfix}
\usepackage{xcolor}
\usepackage{multirow}
\usepackage{multicol}
\usepackage{adjustbox}
\usepackage{hyperref}

\makeatletter
\newcommand\footnoteref[1]{\protected@xdef\@thefnmark{\ref{#1}}\@footnotemark}
\makeatother

\begin{document}

\title{Localization Threats in\\Next-Generation Wireless Networks}

\author{Caner~Goztepe, Saliha~Buyukcorak, Gunes~Karabulut~Kurt,~\IEEEmembership{Senior~Member,~IEEE}, Halim~Yanikomeroglu,~\IEEEmembership{Fellow,~IEEE}

\thanks{C. Goztepe is with the Department of Electronics and Communications Engineering, Istanbul Technical University, Istanbul, Turkey, 30332 (e-mail: goztepe@itu.edu.tr)}
\thanks{S. Buyukcorak is with the Department of Electronics Engineering, Gebze Technical University, Kocaeli, Turkey, 41400 (e-mail: sbuyukcorak@gtu.edu.tr) }
\thanks{G. Karabulut Kurt was with the Department of Electronics and Communications Engineering, Istanbul Technical University, when this work was performed. She is now with the Poly-Grames Research Center, Department of Electrical Engineering, Polytechnique Montr\'eal, Montr\'eal, QC, Canada (email: gunes.kurt@polymtl.ca). }
\thanks{H. Yanikomeroglu is with the Department of Systems and Computer Engineering, Carleton University, Ottawa, Ontario, Canada (e-mail: halim@sce.carleton.ca)}
}
\maketitle

\begin{abstract}
The impact of localization systems in our daily lives is increasing. As next-generation networks will introduce hyper-connectivity with the emerging applications, this impact will {undoubtedly} further increase, proliferating the importance of the location information's reliability. As society becomes more dependent on this information in terms of the products and services, security solutions will have to be enriched to provide countermeasures sufficiently advanced to ever-evolving threats, forcing the joint {design} of communication and localization systems. This paper {envisions} integrated communication and localization systems by focusing on localization security{. Also,} conventional and next-generation attacks on localization are {discussed} along with an {efficient attack} detection method and test-bed{-}based demonstration, highlighting the need for effective countermeasures.
\end{abstract}

\vspace{-0.5cm}
\section{Introduction}

\begin{figure*}[!ht]
\centering
\includegraphics[width=0.85\linewidth]{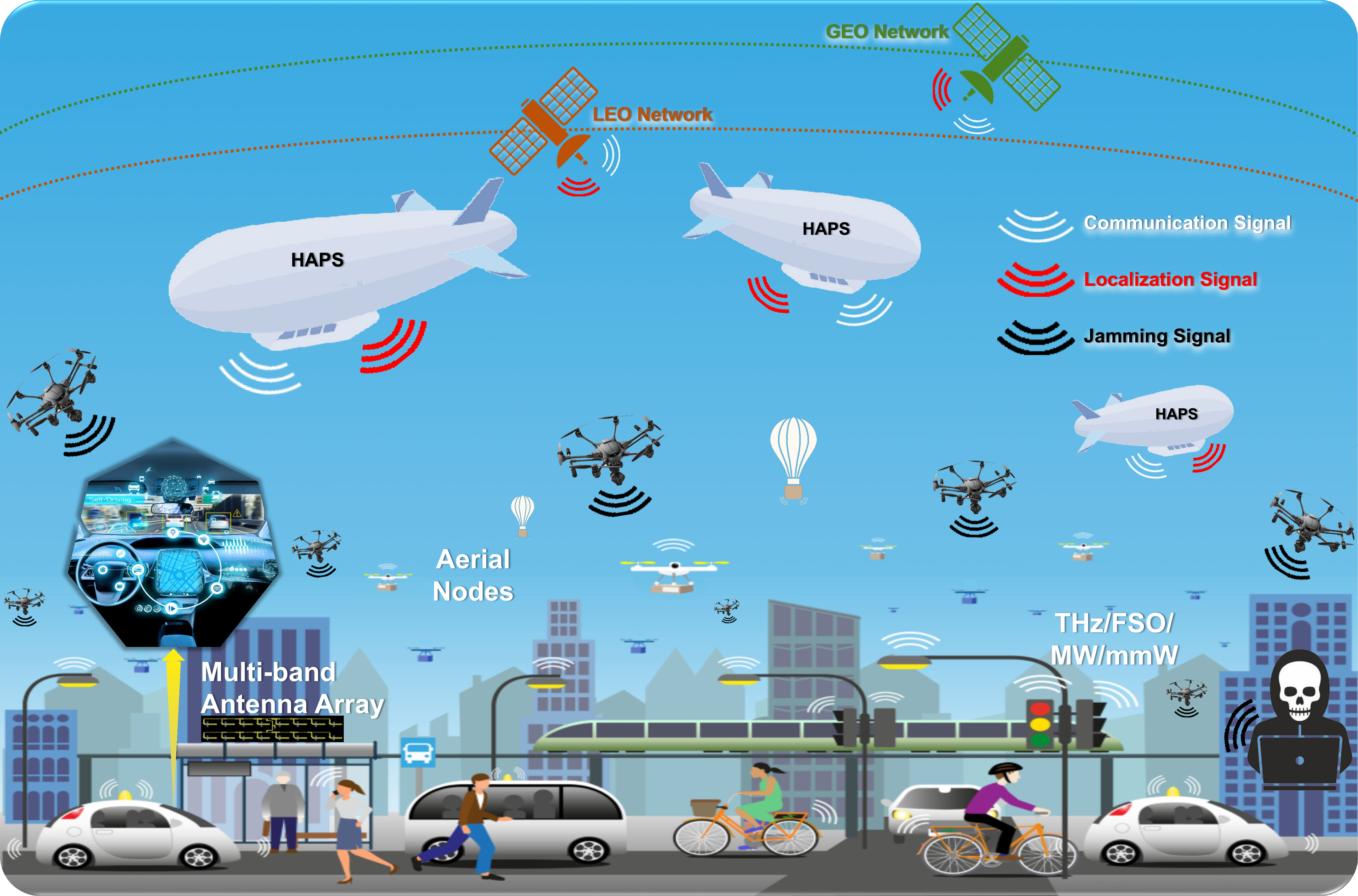}
\caption{A vision of ICLS on the research path towards 6G with the emerging architectural technologies and user applications.}
\label{fig:app}
\end{figure*}

{The p}rogress in user companion devices equipped with rich communication capabilities and advanced computational intelligence {forces} wireless networks to enable the emerging computation-hungry and delay-sensitive services. These trends, in addition to revealing more stringent communication requirements in terms of rate-reliability-latency, also trigger precise and highly accurate secure localization systems. From the product and services perspectives, the research path towards 6G dauntingly forces {the} design of \textit{integrated wireless communication and localization systems} (ICLS), as visualized in Fig.~\ref{fig:app}. 

{The r}esearch community has been working on localization systems with a mission-critical part of various civilian and military applications for decades. Current well-established solutions include GNSS-based localization for outdoors and WiFi/Bluetooth\textbf{-}based localization for indoors. These solutions provide a limited accuracy, depending on the environment and products 3-10 meters (m) resolution for GNSS and 0.3-2.5 m for WiFi/Bluetooth \cite{IndLocSURVEY,LocalizationBOOK}.
Nevertheless, localization remains an active research area {regarding} the achieved performances and the practical implementations, particularly in the emerging new usage applications that desire cm-level (even below) localization accuracy \cite{THz}.

More importantly, the existing localization solutions mainly focus on localization performance{,} assuming reliable and cooperative signal sources, regardless of localization security. However, they are also vulnerable to attacks on localization systems, including jamming attacks {(disrupting} the corresponding localization services{)} and spoofing attacks {(misleading} the target services{)}. Both types of attacks are frequently observed in GNSS-based localization systems \cite{MIT-Review-Paper}. Albeit such malicious attacks to wireless localization systems can cause irreparable damages due to {the} sheer plethora of applications{,} including connected autonomous vehicles, in the literature only a few works concentrate on localization security, such that their countermeasures are founded on just location information perspective \cite{spoof1,jammer}. Wireless localization security will be more {daunting} with the emerging wiser attackers and the disseminating feature{-}proof systems.

The accuracy and the reliability of location information, with the emerging cyber-physical systems {offering exciting new applications, especially in ICLS,} will be even more critical. 
Motivated from that, this paper aims to highlight the importance of \textit{localization security}, {which} implies protecting location information and location systems from unauthorized access, use, disclosure, disruption, modification, or destruction to provide integrity, availability, and confidentiality \cite{spoof1}.

\begin{table*}[!b]
\caption{Comparison of {essential} localization solutions.}
\label{tab:compare}
\medskip
\begin{adjustbox}{max width=\textwidth}
\begin{tabular}{c|c|c|c|c||c|c|c|}
\cline{2-8}
 & \multicolumn{4}{c||}{\textbf{Localization Techniques Classifications}} & \multicolumn{3}{c|}{\textbf{Physical Parameters Classifications}} \\ \cline{2-8} 
 & GNSS & Cell-ID & VL & Lidar & RSS & ToA \& TDoA & AoA \\ \hline \hline
\multicolumn{1}{|c||}{Environment} & Outdoor & Outdoor & Indoor & \begin{tabular}[c]{@{}c@{}}Outdoor/Indoor\\(mostly autonomous systems)\end{tabular} & Outdoor/Indoor & Outdoor/Indoor & Outdoor/Indoor \\ \hline
\multicolumn{1}{|c||}{Propagation} & LoS & LoS/NLoS & LoS & LoS & LoS/NLoS & LoS & LoS \\ \hline
\multicolumn{1}{|c||}{Accuracy Level} & Medium & Low & High & High & Medium & Medium/High & Medium/High \\ \hline 
\multicolumn{1}{|c||}{Synchronization} & Required & Not Required & \begin{tabular}[c]{@{}c@{}} Not Required\\(Except for ToA and AoA)\end{tabular} & Not Required & Not Required & \begin{tabular}[c]{@{}c@{}}Required\\(not required for two-way-ranging)\end{tabular} & Required \\ \hline
\multicolumn{1}{|c||}{Antenna Array} & Not Required & Not Required & \begin{tabular}[c]{@{}c@{}} Not Required\\(Except for AoA)\end{tabular} & Not Required & Not Required & Not Required & Required \\ \hline
\multicolumn{1}{|c||}{Attack Resilience} & Fairly Low & Medium & Medium & Medium & Low & Medium & Low \\ \hline
\multicolumn{1}{|c||}{Algorithms} & Trilateration & Proximity & Mostly Least Square & - & \begin{tabular}[c]{@{}c@{}}Basically Trilateration \\ Fingerprinting and Others* \end{tabular} & Trilateration & Trilateration \\ \hline 
\end{tabular}
\end{adjustbox}
* {\tiny Maximum likelihood estimators (Multi-resolution scaling and Optimization techniques), Bayesian approaches (Expectation-Maximization, Metropolis-Hastings), Machine learning approaches}
\end{table*}

Our five contributions are {as follows:}
\begin{enumerate} [{C}1]
\item We give a crisp literature review on the existing wireless localization systems, {focusing} on the attacks targeting these systems, as summarized in Table \ref{tab:compare} (Section \ref{sec:methods}).

\item We {provide a vision for} ICLS, especially by focusing on wireless localization systems, along with providing a vision of the leading applications and the architectural changes that are emerged in the next-generation networks towards 6G (Section \ref{sec:applications}). 

\item We describe conventional jamming and spoofing attacks against ICLS and introduce a {detection} concept for such attacks based on verifying integrated frame components' information content. 
We {initiate} possible next-generation attack concept in ICLS and propose {a practical} {Kullback–Leibler (KL) divergence based detection method} (Section \ref{sec:attack}).

\item We design a real-time software-defined radio (SDR) based test-bed along with conventional and next-generation attack concepts and quantify the impact of these attacks on ICLS {through the comprehensive measurements} (Section \ref{sec:measurements}).

\item We discuss open research problems that need to be addressed by the research community (Section \ref{sec:OI}). 
\end{enumerate}

\section{Overview of Wireless Localization Systems}\label{sec:methods}

\subsection{Classifications of Physical Parameters}
The parameters {of localization systems}, include received signal strength (RSS), time (difference) of arrival (T(D)oA), angle of arrival (AoA) {depending on the available hardware}. These are {summarized} up in Table \ref{tab:compare}.

\textit{RSS:} 
It is {affected} by three multiplicative factors: path loss, shadowing and fading. As fading in the localization solutions is eliminated by averaging the measured signal levels, typically, it is characterized by {a} log-distance path loss model (as a function of the distance between the target and the anchor, path loss exponent{,} and transmit power) and lognormal shadowing \cite{SURVEY_WLAN}.

\textit{T(D)oA:}
ToA {captures} the distances between the target and the anchors from the propagation time measurements assuming that the signal's velocity is known. It can be grouped into one-way-ranging (requires precise synchronization) and two-way-ranging (does not require synchronization). TDoA, a variation of ToA, provides the differences among these distances and needs perfect synchronization between the anchors.

\textit{AoA:}
The distance is extracted from the angle between the target and the anchor {by concerning} a reference direction. It ensures high localization accuracy (down to a few degrees). However, the anchor requires the adoption of antenna arrays and a minimum distance between the antenna elements {to estimate the angle} \cite{IndLocSURVEY}.

\vspace{-0.2cm}
\subsection{Classifications of Localization Techniques}

The available localization systems can be grouped into two categories: global localization systems and local localization systems \cite{LocalizationBOOK}. Local systems, viewed as relative localization systems, can be divided into self-localization systems ({where} the target finds its location with respect to anchors with known location) and remote localization systems ({where} each device within a coverage area finds the relative location of other devices in that area). The leading localization techniques are briefed in Table \ref{tab:compare}.

\textit{GNSS-Based Localization:} 
It is a general term to define any satellite constellation for ensuring autonomous geo-spatial localization, navigation, and timing services on a global or regional basis. There exist different satellite systems, including GPS (US), BeiDou (China), Galileo (EU), GLONASS (Russia), IRNSS (India), and QZSS (Japan) \cite{LocalizationBOOK}.

\textit{Cell-ID-Based Localization:}
A Cell-ID is a unique key, generally {a} number, which {can be used} to identify each base transceiver station or its sector within a location area code. It ensures the identity or geographical description of the cell to which the target is connected. Cell-ID, due to its {uniqueness} as well as simplicity and low cost, is the preferable method for target localization in cellular communication systems. It provides low localization accuracy even for outdoors \cite{SURVEY_CELLULAR}.

\textit{Visible Light (VL) based Localization:}
VL has been a powerful, promising technology for accurate and reliable indoor localization, resulting from that (i) LED sources to enable economical and efficient lighting as well as communication-based on the imperceptible flickering light signals, (ii) the signals are not affected by RF electromagnetic interference \cite{VLC}. However, this technique is mainly useful at short ranges due to the propagation characteristics of the light signals.

\textit{Ra(Li)dar:} 
Radar renders the distances to environmental features by emitting radio signals and detecting reflected signals \cite{LIDAR}. It exploits the round-trip time of the signals between the radar and the target. Lidar originates from the enforcement of laser light signals to radar systems. Main Lidar techniques for measuring the distance are the pulse measurement and the phase shift measurement \cite{LIDAR}. Radar and Lidar solutions have {been} actively {used} in autonomous driving control and navigation recently.

\vspace{-0.32cm}
\subsection{Classifications of Localization Algorithms}

\textit{Trilateration:}
Localization is made with the intersection of geometric forms, such as circles, triangles, or hyperbolas, created by {measuring} distances from multiple anchors. 

\textit{Fingerprinting:}
This technique, being a scene analysis, is frequently utilized in RSS-based localization {systems}. It operates in two primary phases. The offline training phase builds a fingerprint measurements map through the anchors with known locations on the monitored region.
The online localization phase determines the unknown target location by exploiting fingerprint measurements (using algorithms such as the nearest neighbor (NN), KNN, and weighted KNN).

\textit{Maximum Likelihood Estimators:}
Due to its asymptotic optimality, {they have been widely used} in practice. Unfortunately, the nature of localization problems prohibits obtaining a closed-form solution. Numerical methods, including multidimensional scaling and optimization techniques, can be used to {obtain the solution}. The drawback{s} of {numerical methods are} {time-consumtion} and {high} memory requirement. {It} is difficult to transform the original nonconvex and nonlinear problem into a convex problem{, when optimization techniques are used.}

\textit{Bayesian Approaches:} 
These are of great importance for highly {accurate} and reliable localization techniques to ensure more robustness against varying channel propagation conditions and measurement uncertainties. They also perform {well} in highly dynamic and adaptive next-generation networks.

\vspace{-0.28cm}
\section{Toward Next-Generation Networks} \label{sec:applications}

{E}merging rate-reliability-latency critical services and user applications, as well as the dramatic architectural changes towards next-generation networks, force the design of ICLS instead of the well-studied separate communication and localization systems. 
This paper on the research path towards 6G envisages ICLS, as illustrated in Fig. \ref{fig:app}. 
This figure provides a vision of the leading applications and the architectural changes, as detailed below, along with possible attacks, especially by focusing on wireless localization systems. 

\vspace{-0.4cm}
\subsection{Emerging ICLS Applications} 

5G and the envisioned next-generation 6G networks aim to serve as {enablers to} Internet of Everything (IoE) networks. These networks are expected to connect billions of machines and millions of people and introduce innovative applications that address high rate and low latency constraints. However, in addition to the mentioned connectivity constraints, highly accurate and secure location (possibly in 3D) information needs to be provided to be fully immersed in emerging IoE applications. 

Multi-sensory extended reality (XR) applications, including augmented, mixed, and virtual reality (AR/MR/VR), are {expected to be the essential} applications of IoE \cite{6G}.
Immersive XR applications are expected to project cognitive capabilities according to visual and haptic perceptions. To fully reflect human perception devices, communication, computing, and storage capabilities need to be supported by accurate and reliable location information, possibly in cm-level, that work indoors and outdoors.

Another promising application of IoE is autonomous systems. {H}ighly complex autonomous systems must identify the suitable action or set of actions under the current conditions {according to the target objectives}. If actual conditions and perceived conditions are not matching, a series of erroneous actions with severe consequences may occur. 
Exemplary systems include autonomous cars and crewless autonomous vehicles. Such systems can be mobile in 2D, such as platooning trucks, or 3D, such as cargo-drones and drone swarms. The integrated control process associated with the autonomous systems introduces the requirement of accurate location information of both the corresponding device and the surrounding devices.

Industrial applications of IoE are also expected to play a significant role in the manufacturing industry of the near future. Internet of Things (IoT) and Industrial IoT networks are anticipated to evolve towards industrial IoE
, where fully automated factories become a reality that is supported through cognitive functionalities within the manufacturing processes. In these scenarios, accurate and secure location information in indoor deployments is {also} a must.

\subsection{Architectural Changes}
\vspace{-0.15cm}
\textit{Integrated Aerial Terrestrial Networks:}
{W}ireless networks are evolving to accommodate aerial networking elements capable of moving in 3D both as the base stations (BS) and the user equipment. As suggested in 3GPP TS 22.125 of Release 17\footnote{\label{note1}{Available at \url{https://portal.3gpp.org/}}}, terrestrial networks will be supported through drone-mounted BS that fly up to 200 m. Furthermore, there are intense discussions for support via pseudo-stationary high altitude platform stations (HAPS) allocated for 20km altitude in the atmosphere. Additionally, in the near future, low orbit satellite (LEO) networks will operate harmoniously to the terrestrial networks to satisfy the ever-increasing demand of wireless users. 

This evolution is mainly motivated by the need for agility, high data rate{,} and low latency demands. The use of aerial networking elements implies the 3D mobility of network elements. This ability also enforces accurate and secure location information from both planning and operating perspectives. We note that this 3D network structure can be approximated by 2D if the vertical and horizontal distances are relatively scalable. In terms of planning, the optimal placement of drone BS or their mobile networking elements such as HAPS is a well{-}investigated research topic \cite{drone-placement}. From an operational perspective, the locations of devices are also needed. Drones can operate up to 200 m, so even to simply track the compliance against regulations the location information must be monitored. The necessity for the precise localization requirements in unmanned aerial systems is highlighted in TS 22.261\footnotemark[1], as a horizontal (vertical) accuracy of 0.5 (1) m with 99\% probability.

\textit{Terahertz Band:}
The data rate demands are expected to exceed 100 Gbps. However, current physical layer transmitter and receivers designs fail to address such data rates that require spectral efficiency exceeding 14 bit/s/Hz \cite{THz}. A promising solution to address such data rates is {to use the} THz band, from 100 GHz to {10} THz. This also provides a significant advantage from {the} localization perspective. Due to the use of ultra-wide bandwidth, even finely spaced multi-path components can be accurately resolved. This accurate ToA estimate and accurate AoA estimation and 3D maps can be jointly used for cm-level accuracy.

\textit{Multi-connectivity:} {N}ext-generation wireless localization environments, as visualized in Fig. \ref{fig:app}, become more chaotic and irregular due to increasingly diversified target and anchor nodes. These devices can leverage different wireless networks with several access technologies over multiple frequency bands, and even an aggregation of these networks with multi-connectivity \cite{HETNET}. 
The communication and localization devices in the ICLS could be part of both systems based on signalization characteristics. 
Thus, the integrated system's frame structure should be designed {according to} both communication and localization tasks. 

\begin{figure}[!tb]
\centering
\includegraphics[width=\linewidth]{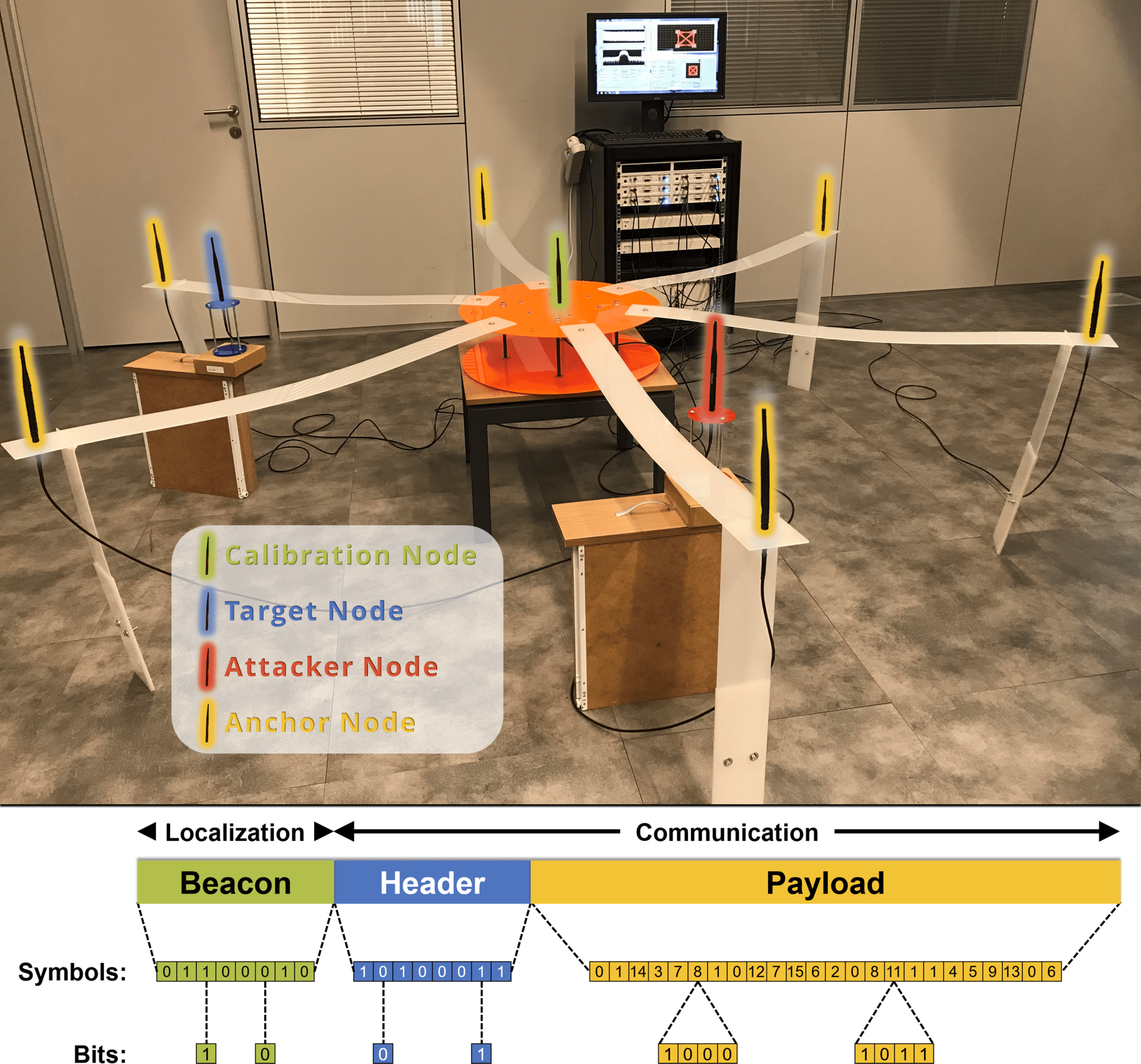}
\caption{SDR-based real-time wireless localization test environment with malicious attacks.}
\label{fig:testbed}
\end{figure}

\section{Attacks to Wireless Localization Systems} \label{sec:attack}

In addition to the {applications} that require much more accurate location information, as smart wireless devices become cheaper and more accessible by malicious users or attackers, localization systems get more vulnerable to attacks misleading the target{'s} location or communication information.

\begin{figure*}[!tb]
\centering
\includegraphics[width=.95\linewidth]{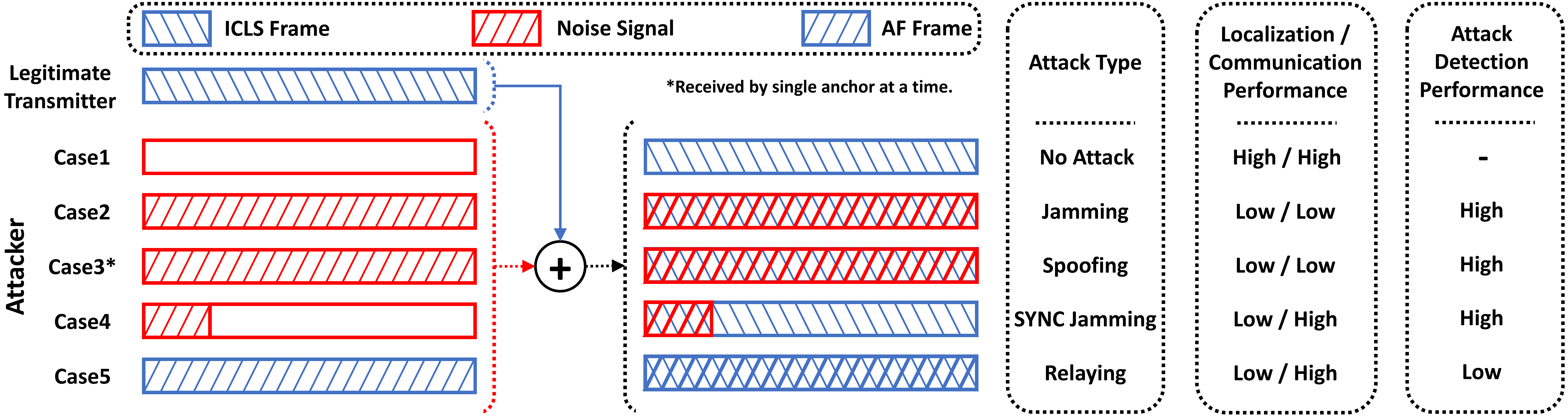}
\caption{Description and comparison of localization attack cases on ICLS.}
\label{fig:cases}
\end{figure*}

\vspace{-0.25cm}
\subsection{Conventional Attacks}
Conventional attacks on localization systems are composed of jamming and spoofing attacks. \textit{Jamming attacks} are the attacks where one or more wireless device (jammer) transmit {s} distorting signal to the anchor nodes, frequently transmitted in an omnidirectional manner to cause a varying estimation error with no particular pattern. Collaborative jammer attacks are more challenging to implement due to the coordination needs between {attackers,} yet they are more disruptive \cite{jammer}.
In the presence of multiple jammers, the range and intensity of jamming signal can be higher. {Frequently} used counter-jamming techniques aim to improve the signal quality (e.g., through front-end designs including beamforming) or filtering or resource allocation (e.g., frequency hopping or transmit power increase). Additionally, jammer localization is also considered with the purpose of physically removing the attacker.

\textit{Spoofing attacks} on localization systems aim to intentionally mislead the location information estimated by the anchor nodes \cite{spoof1}. Spoofing attacks {are} studied on a minimal scale in the literature, yet their impact is already costly in the GNSS localization. In the presence of emerging applications, such as autonomous vehicles, spoofing attacks that randomly distort the location estimate but misleads the system can even cause physical accidents. Finding anti-spoofing solutions is an indispensable requirement. Frequently used anti-spoofing techniques include monitoring the power levels, either the average power or its distribution. Anomaly detectors are commonly utilized {for} this purpose to determine the presence of a spoofer.

{The a}ttacker's transmission is {independent from the transmissions} of the communication system in conventional attacks. The anchor nodes can detect such attacks by jointly monitoring the location estimator and the receiver. {This feature can} be enabled by the ICLS. Physical layer characteristics of the received signals, such as statistical information about RSS and bit error rates (BER), render high accuracy detection of these attacks, as demonstrated in Section \ref{sec:measurements}. Such attacks can also be detected by merging the localization system's finding with the authentication processes.

The conventional attacks do not vary their attack model as per the transmission state of the target. Threats to localization systems are expected to increase as the attackers' abilities expand and {they can synchronize or} couple themselves with the ongoing packet transmissions. This would allow the attackers to change their transmitted signals following the target's transmission status, as detailed below.

\vspace{-0.1cm}
\subsection{Next-Generation Attacks} 

Referred here as the next-generation attacks, we need to highlight the emerging threat by possible coupling between the ICLS and the localization attacker. With more apriori information about the signaling, {the attacker's transmission can be coupled with the transmission of the communication system, and the attacker can tune} its transmission by the legitimate system, renders itself to be undetectable, as in covert communications scenarios. As attackers can {modify} their transmission {according to} that of the target, the attacks' plethora can dramatically expand. For instance, a protocol-aware attacker can attack solely the beacon, or header of the frame structure, while not targeting the payload. Here, we give two examples of such attacks. To provide concrete attack examples, we consider a three-component frame structure, as {shown} in Fig. \ref{fig:testbed}. These are the data payload (ensures data communication), the header (includes address and protocol-related information that can be used authentication process), and the preamble (procures synchronization and timing estimation) \cite{Frame}. The preamble can also be utilized as a beacon component for localization and can support communication if necessary.

In the first type of attack, the attacker can synchronize its signals with that of the ICLS, as shown in Case 4 {of} Fig. \ref{fig:cases}. This allow{s} the attacker to transmit only through certain parts of the ICLS frame used for localization. Since the attack is intermittent in time, the probability of detection is reduced when averaging-based detection techniques are used. Nevertheless, it is possible to separately monitor the statistics about the ICLS's all frame components to detect these attacks. Such an attack on the beacon signal is demonstrated in Section V, where the attacker only transmits during the {beacon's transmission}. In {the} case of monitoring the transmission characteristics according to the frame partitions, the average power differences between the received beacon and the received header help reveal this attack's presence. Note that if the communication receiver and the location estimator are not integrated, this attack can not be detected.

\begin{figure*}[ht!]
\centering
	\begin{subfigure}{0.45\textwidth}
		\centering
		\includegraphics[height=5.7cm]{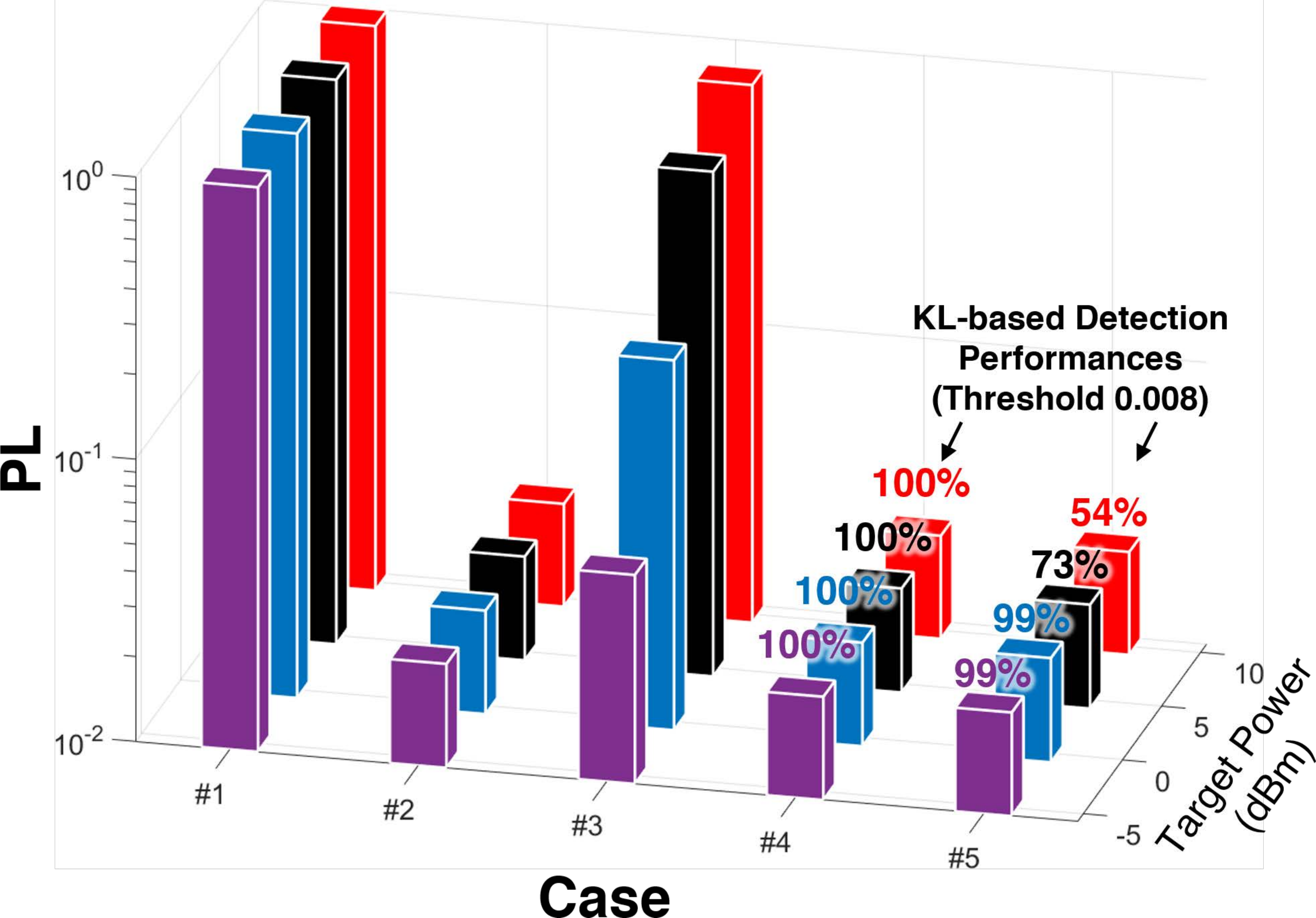}
		\caption{ }
		\label{fig:pcdresults}
	\end{subfigure}
	\begin{subfigure}{0.45\textwidth}
		\centering
		\includegraphics[height=5.7cm]{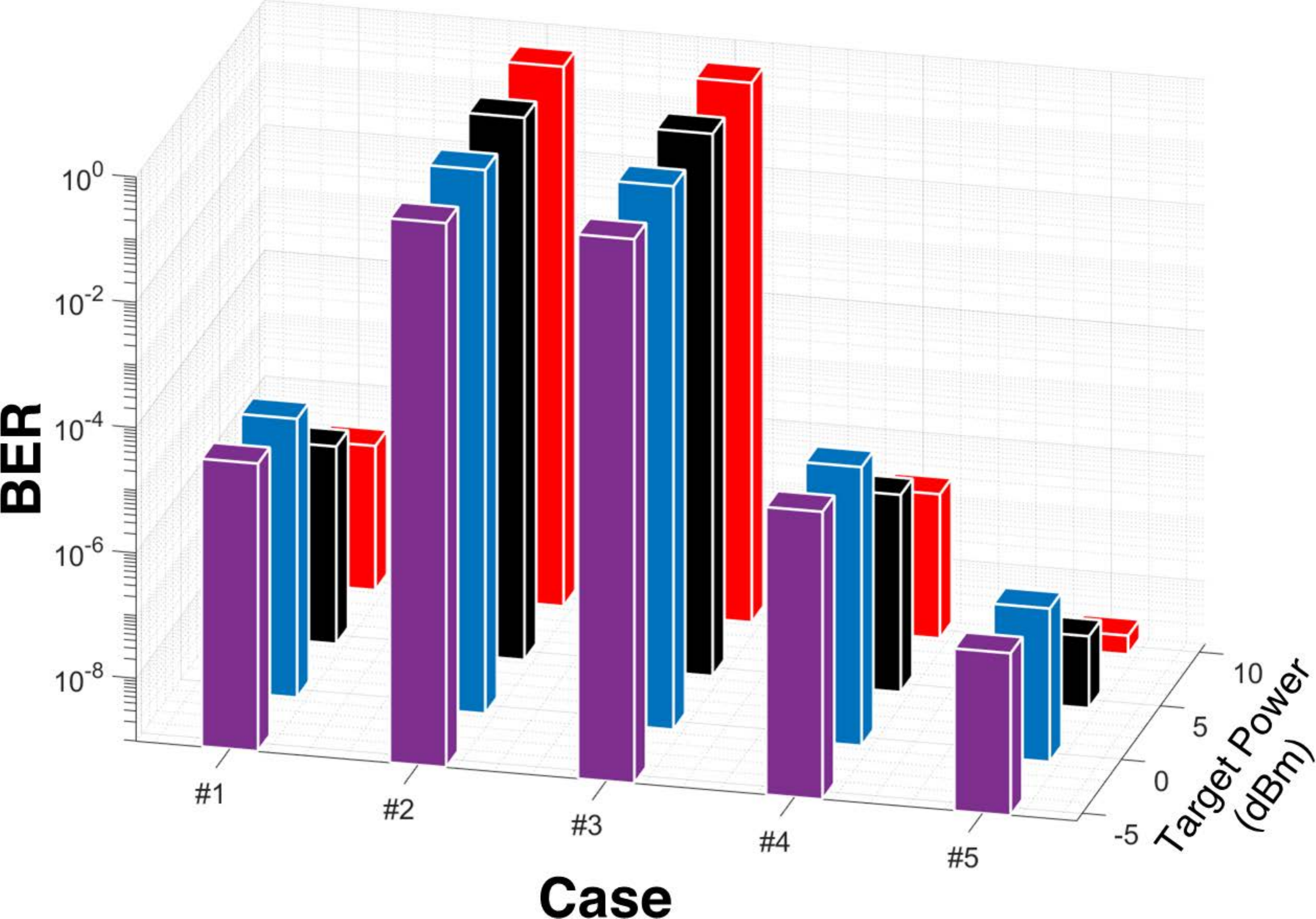}
		\caption{ }
		\label{fig:berresults}
	\end{subfigure}
	\vspace{-0.05cm}
	\caption{(a) The measurement results for the probability of accurately finding the unknown target location. (b) The error measurement results {in} the communication data.}
	\label{fig:results}
\end{figure*}

As the second kind of attack, we consider a tighter coupling between the attacker and ICLS. In this attack, shown as Case 5 {of} Fig. \ref{fig:cases}, 
the attacker aims to avoid detection even when a frame-partition based detection technique is used. With this purpose, the attack is continued throughout transmission. To {deceive} the legitimate receiver, the frame content is also kept by using relaying techniques. Hence, we call {it a} relaying attack. The attacker first captures the legitimate signal and then forwards it similar to the man-in-the middle attack yet without changing the content, solely distorting the physical origin of the transmission in a more dominant manner {(}such as with high transmission power{)}. Although both decode-and-forward and amplify-and-forward (AF) techniques can be used, to avoid any processing delay and keep the transceiver structure of the attacker simpler, AF technique is preferred. Note that AF-based signal transmission would still be possible even when the legitimate signals were encrypted. The attacker can forward the authenticated packets through a different transmission path (that emerges naturally when the spoofer and the target node are physically spaced apart). 
Notice that the spoofer does not need the channel information between the target and the anchor nodes. Under this attack, the attacker device's location will be detected instead of the target node's location. A mobile attacker can also make that the spoofer change the erroneous location information depending on the attack{'}s motives. 
The average powers and statistical information about frames do not reveal a significant difference as they are merely forwarded by the attacker. The performance of the communication system is not affected by this kind of attack. 

The presence of double fading channel characteristics for relaying attack can reveal a hint about this attack's presence, hence this paper proposes an {KL divergence based detection method}.
The proposed method ground on observing the ICLS frames to measure {the} similarity between no attack ({regular} traffic) and attack (anomaly traffic) cases through {the} KL metric. 
It decides whether there is an attack by comparing the KL divergence values with a static value as {a} threshold.

As the SDR units become cheaper, more capable, and more accessible by the attackers, more attacks may be encountered. In the following section, by demonstrating the applicability of these attacks, we warn about this imminent threat that needs to be addressed urgently.

\vspace{-0.15cm}
\section{Measurement Results} \label{sec:measurements}

This section, through the designed SDR-based real-time ICLS test-bed (has been reduced to 2D due to the scalability of vertical and horizontal distances) for conventional and next-generation attacks, demonstrates the vulnerabilities of wireless localization systems to malicious attacks, and the effectiveness of the proposed {KL divergence based detection} method.
Five different transmission cases are considered. In the first case, no attacker is present. In the second and third cases, the jamming and spoofing attacks are performed. The two examples given in Section IV.B are the fourth and the fifth attacks, respectively. An overview of the attacks is given in Fig. \ref{fig:cases}.

Accordingly, the test-bed, as illustrated in Fig. \ref{fig:testbed}, consists of 9 nodes on 5 USRP-2943Rs, each with two RF chains. A 6-cell neighborhood is constructed by the six anchor nodes representing a cell. The antenna at the center is used for calibration and synchronization purposes. The attacker node supports full-duplex communication. All nodes are controlled by the same host computer with LabVIEW program. The anchors collect the transmitted frame shown in Fig. \ref{fig:testbed}, which includes BPSK modulated Beacon (8 bits), BPSK modulated Header (8 bits), and 16-QAM modulated Payload (96 bits) with 2.45 GHz frequency band and 1 MHz bandwidth. The unknown target location is estimated using beacon's IQ measurement values with grid-search-based localization by considering the six possible positions.

The real-time test measurements are performed with the target powers varying as ($-5, 0, 5, 10$) dBm to consider possible received power values ranging from lower to higher, and the attacker power is $20$ dBm that is the maximum output power the devices can provide. For the listed attacks in Fig. \ref{fig:cases} on {the} ICLS, the localization performance is quantified with {probability of localization (PL) that refers to selecting} the correct cell. Header BER performance is measured and used jointly with PL for the attack detection.

An overview of the results is given in Fig. \ref{fig:cases}, and Fig. \ref{fig:results} demonstrates the real-time test results. 
We observe that the target accurate location information cannot be determined for relaying attack (Case 5), and PL is approximately 18\%. This measurement work reveals that cyber-physical attacks, easily implemented, grievously disrupt wireless localization systems' performance, and countermeasures {must} be immediately addressed. We also see from Fig. \ref{fig:pcdresults} for Case 4 and {Case} 5 that the proposed {attack detection} method achieves {a} 100\% detection rate when the threshold value is 0.008.


{Our test-bed consists of SDR nodes that can be used for multiple purposes. Such that these nodes can be operated as a multi-functional receiver/transmitter via software control, differ from a single-function receiver/transmitter structure. Hence, this test-bed is a preliminary preparation for multi-functional receiver and transmitter structures to be practiced in the near future. It can also be included a countermeasure to preserve against any attack, as well as attack detection. Apart from testing the proposed ICLS with several parameters, the test-bed can satisfy many different system suggestions with its multi-antenna structure. Additionally, the machine learning algorithms can be efficiently operated on this test-bed through its software configurable nature.}

\vspace{-0.15cm}
\section{Open Issues} \label{sec:OI}
 
\textit{Effective Countermeasures against Localization Attacks:}
As the attacks on wireless localization system that will be shaping our daily activities very soon get more sophisticated, effective defense strategies become critical than ever before. The current single-{layer} solutions will fail to detect and react to these attacks. To improve the efficacy defense mechanisms, information from the physical layer, data layer, network layer, and even application layer can be combined, paving the way for cross-layer aided approaches. Decision fusion systems can be used to enhance the resistance against localization attacks. ICLS must also be designed to protect the privacy of the users.

\textit{Sub-mm Localization Applications:}
Increasing localization accuracy will pave the way for further applications over cyber-physical systems. Sub-mm accuracy will be able to trigger a diverse range of biomedical applications. For instance, such high accuracy localization systems can enable the implementation of microfluidic systems for {the} miniaturization of chemical and biological processes. The applications offered by the micro-electromechanical devices can further be diversified with localization support. Micro/nano-robots equipped with localization sensors can reveal in vivo tracking. These applications, fueled by imagination, can offer life-changing benefits in the near future, further increasing the security and accuracy requirements on ICLS.

\textit{Localization of Aerial Network Nodes:}
The integrated aerial terrestrial networks will introduce the mobility challenges of the localization systems. Locations of aerial networking components need to be available against security attacks. Considering the dense LEO constellations to be deployed in the near future within the plans of SpaceX, {and} Amazon with Project Kuiper, the Doppler shifts and propagation delay variation may reach 55 kHz and 20 $\mu$s/s, respectively \cite{sat-network}. Drone-mounted aerial BS or cargo drones that are connected to the wireless networks as user equipment nodes also have high mobility capability, along with (secure) localization requirements. Continuous tracking of these mobile nodes is also necessary and must be enabled. Obviously, the spoofed or jammed location information of these devices may create high risk in network management.

\textit{Emerging Machine Learning Approaches:}
Its use in wireless localization systems has been extensively investigated for over three decades. Nevertheless, the emerging and much-needed explainable and human-understandable artificial intelligence techniques, such as explainable deep-learning and federated-learning approaches, can be a remedy for ICLS, where model-based approaches fail to provide a sufficient localization accuracy. These techniques will also provide more transparency against jamming and spoofing threats on the localization systems.

\textit{Frame Design:} This should be considered under the required communication and localization performance. Mimicking the echolocation signals of bats and whales, for the preamble, the efficacy of the low peak-to-average-power chirp signals is a prominent waveform candidate. The literature on radar signals can also be combined from this perspective to improve the overall system performance.

\vspace{-0.4cm}
\section{Conclusions} \label{sec:conc}

{L}ocation information is a valuable context, and any attack on this information has critical consequences. {J}ammed or spoofed location{s} may have more severe consequences. We provide an overview of the attacks on wireless localization systems, mainly focusing on the conventional and the next-generation attacks. We also propose an efficient {Kullback–Leibler divergence based detection method.}
We demonstrate the threat by deploying an exemplary mini-scale localization system {using software-defined radio nodes}.

\bibliographystyle{IEEEtran}
\bibliography{ref}

\begin{thebibliography}{10}
\providecommand{\url}[1]{#1}
\csname url@samestyle\endcsname
\providecommand{\newblock}{\relax}
\providecommand{\bibinfo}[2]{#2}
\providecommand{\BIBentrySTDinterwordspacing}{\spaceskip=0pt\relax}
\providecommand{\BIBentryALTinterwordstretchfactor}{4}
\providecommand{\BIBentryALTinterwordspacing}{\spaceskip=\fontdimen2\font plus
\BIBentryALTinterwordstretchfactor\fontdimen3\font minus
  \fontdimen4\font\relax}
\providecommand{\BIBforeignlanguage}[2]{{%
\expandafter\ifx\csname l@#1\endcsname\relax
\typeout{** WARNING: IEEEtran.bst: No hyphenation pattern has been}%
\typeout{** loaded for the language `#1'. Using the pattern for}%
\typeout{** the default language instead.}%
\else
\language=\csname l@#1\endcsname
\fi
#2}}
\providecommand{\BIBdecl}{\relax}
\BIBdecl

\bibitem{IndLocSURVEY}
F.~{Zafari}, A.~{Gkelias}, and K.~K. {Leung}, ``A survey of indoor localization
  systems and technologies,'' \emph{IEEE Commun. Surveys Tuts.}, vol.~21,
  no.~3, pp. 2568--2599, Thirdquarter 2019.

\bibitem{LocalizationBOOK}
S.~A.~R. Zekavat and R.~M. Buehrer, \emph{Handbook of Position Location}.\hskip
  1em plus 0.5em minus 0.4em\relax Hoboken, New Jersey: John Wiley \& Sons,
  Inc., 2019.

\bibitem{THz}
T.~S. {Rappaport}, Y.~{Xing}, O.~{Kanhere}, S.~{Ju}, A.~{Madanayake},
  S.~{Mandal}, A.~{Alkhateeb}, and G.~C. {Trichopoulos}, ``Wireless
  communications and applications above 100 {GHz: Opportunities} and challenges
  for {6G} and beyond,'' \emph{IEEE Access}, vol.~7, pp. 78\,729--78\,757,
  2019.

\bibitem{MIT-Review-Paper}
M.~Harris, ``Ghost ships, crop circles, and soft gold: A {GPS} mystery in
  {S}hanghai,'' MIT Technology Review, Tech. Rep., 2019.

\bibitem{spoof1}
J.~H. {Lee} and R.~M. {Buehrer}, ``Location spoofing attack detection in
  wireless networks,'' in \emph{IEEE GLOBECOM}, Dec. 2010, pp. 1--6.

\bibitem{jammer}
S.~{Gezici}, M.~R. {Gholami}, S.~{Bayram}, and M.~{Jansson}, ``Jamming of
  wireless localization systems,'' \emph{IEEE Trans. Commun.}, vol.~64, no.~6,
  pp. 2660--2676, Jun. 2016.

\bibitem{SURVEY_WLAN}
A.~{Khalajmehrabadi}, N.~{Gatsis}, and D.~{Akopian}, ``Modern {WLAN}
  fingerprinting indoor positioning methods and deployment challenges,''
  \emph{IEEE Commun. Surveys Tuts.}, vol.~19, no.~3, pp. 1974--2002,
  Thirdquarter 2017.

\bibitem{SURVEY_CELLULAR}
J.~A. {del Peral-Rosado}, R.~{Raulefs}, J.~A. {López-Salcedo}, and
  G.~{Seco-Granados}, ``Survey of cellular mobile radio localization methods:
  From {1G} to {5G},'' \emph{IEEE Commun. Surveys Tuts.}, vol.~20, no.~2, pp.
  1124--1148, Secondquarter 2018.

\bibitem{VLC}
S.~{Büyükçorak} and G.~{Karabulut Kurt}, ``A {Bayesian} perspective on {RSS}
  based localization for visible light communication with heterogeneous
  networks extension,'' \emph{IEEE Access}, vol.~5, pp. 17\,487--17\,500, 2017.

\bibitem{LIDAR}
A.~{Eskandarian}, C.~{Wu}, and C.~{Sun}, ``Research advances and challenges of
  autonomous and connected ground vehicles,'' \emph{IEEE Trans. Intell. Transp.
  Syst.}, pp. 1--29, 2019.

\bibitem{6G}
W.~{Saad}, M.~{Bennis}, and M.~{Chen}, ``A vision of {6G} wireless systems:
  {Applications,} trends, technologies, and open research problems,''
  \emph{IEEE Network}, vol.~34, no.~3, pp. 134--142, Jun. 2019.

\bibitem{drone-placement}
R.~I. {Bor-Yaliniz}, A.~{El-Keyi}, and H.~{Yanikomeroglu}, ``Efficient {3-D}
  placement of an aerial base station in next generation cellular networks,''
  in \emph{IEEE ICC}, May 2016, pp. 1--5.

\bibitem{HETNET}
S.~{Büyükçorak}, G.~K. {Kurt}, and A.~{Yongaçoğlu}, ``Inter-network
  localization frameworks for heterogeneous networks with multi-connectivity,''
  \emph{IEEE Trans. Veh. Technol.}, vol.~68, no.~2, pp. 1839--1851, Feb. 2019.

\bibitem{Frame}
J.~Lindh, ``Bluetooth low energy beacons,'' Texas Instruments, Tech. Rep.,
  2016.

\bibitem{sat-network}
W.~{Wang}, Y.~{Tong}, L.~{Li}, A.~{Lu}, L.~{You}, and X.~{Gao}, ``Near optimal
  timing and frequency offset estimation for {5G} integrated {LEO} satellite
  communication system,'' \emph{IEEE Access}, vol.~7, pp. 113\,298--113\,310,
  2019.

\end{thebibliography}

\vspace{-0.4cm}
\section*{Biographies}
\footnotesize{

CANER GOZTEPE (goztepe@itu.edu.tr) received the B.Sc. and M.Sc. degrees in telecommunication engineering from Istanbul Technical University (ITU), Turkey, in 2017 and 2019. He is currently studying for a Ph.D. degree at ITU. His research interests are 5G+, physical layer security, and SDR applications beyond 5G physical layer schemes.\\

\vspace{-0.05cm}
SALIHA BUYUKCORAK (sbuyukcorak@gtu.edu.tr) received the B.Sc., and Ph.D. degrees in tele}communication engineering from ITU, Turkey, in 2011, and 2019, respectively. She is currently an assistant professor with the Department of Electronics Engineering, Gebze Technical University, Gebze, Turkey. Her research interests include wireless localization systems, wireless channel models, and (non-)terrestrial networks.\\

\vspace{-0.05cm}
GUNES KARABULUT KURT [StM'00, M'06, SM'15] (gkurt@itu.edu.tr) received a Ph.D. degree in electrical engineering from the University of Ottawa, Ottawa, ON, Canada, in 2006. She is now with the Department of Electrical Engineering, Polytechnique Montr\'eal. She is serving as an Associate Technical Editor of IEEE Communications Magazine.\\

\vspace{-0.05cm}
HALIM YANIKOMEROGLU [F] (halim@sce.carleton.ca) is a full professor in the Department of Systems and Computer Engineering at Carleton University, Ottawa, Canada. His research interests cover many aspects of 5G/5G+ wireless networks. His collaborative research with the industry has resulted in 37 granted patents. He is a Fellow of the Engineering Institute of Canada and the Canadian Academy of Engineering. He is a Distinguished Speaker for IEEE Communications Society and IEEE Vehicular Technology Society.

\end{document}